# All-2D Material Inkjet-Printed Capacitors: Towards Fully-Printed Integrated Circuits

Robyn Worsley[1], Lorenzo Pimpolari[2], Daryl McManus[1], Ning Ge[3], Robert Ionescu[3], Jarrid A. Wittkopf[3], Adriana Alieva[1], Giovanni Basso[2], Massimo Macucci[2], Giuseppe Iannaccone[2], Kostya S. Novoselov[4], Helen Holder[3], Gianluca Fiori[2] and Cinzia Casiraghi[1*]

1 School of Chemistry, University of Manchester, Manchester, M13 9PL, UK
2 Dipartimento di Ingegneria dell'Informazione, Università di Pisa, Pisa, 56122, Italy.
3 HP Labs, 1501 Page Mill Road, Palo Alto, California 94304, USA
4 School of Physics and Astronomy, University of Manchester, Manchester M13 9PL, UK

*email:cinzia.casiraghi@manchester.ac.uk

**Abstract**

A well-defined insulating layer is of primary importance in the fabrication of passive (*e.g.* capacitors) and active (*e.g.* transistors) components in integrated circuits. One of the most widely known 2-Dimensional (2D) dielectric materials is hexagonal boron nitride (hBN). Solution-based techniques are cost-effective and allow simple methods to be used for device fabrication. In particular, inkjet printing is a low-cost, non-contact approach, which also allows for device design flexibility, produces no material wastage and offers compatibility with almost any surface of interest, including flexible substrates.

In this work we use water-based and biocompatible graphene and hBN inks to fabricate all-2D material and inkjet-printed capacitors. We demonstrate an areal capacitance of $2.0 \pm 0.3$ nF cm$^{-2}$ for a dielectric thickness of ~3 µm and negligible leakage currents, averaged across more than 100 devices. This gives rise to a derived dielectric constant of $6.1 \pm 1.7$. The inkjet printed hBN dielectric has a breakdown field of $1.9 \pm 0.3$ MV cm$^{-1}$. Fully printed capacitors with sub-µm hBN layer thicknesses have also been demonstrated. The capacitors are then exploited in two fully printed demonstrators: a resistor-capacitor (RC) low-pass filter and a graphene-based field effect transistor.

An integrated circuit (IC) is an assembly of different electronic components, fabricated as a single unit, in which non-linear devices, such as transistors and diodes, and linear devices, such as capacitors and resistors, and their interconnections are all built up on the same substrate, typically silicon.

Two-dimensional (2D) materials[1] are very attractive as building blocks for next-generation nanoelectronic devices as they offer straightforward integration with existing fabrication processes and they can be transferred on to any substrate, including plastic, allowing extending research into flexible electronics, which started with organic materials.[2–8] In addition, the ability to combine different 2D materials in heterostructures[9] allows the possible integration of a wide range of all-2D material devices in ICs. At present, a large family of 2D crystals has been isolated and new 2D materials are constantly being investigated. As such, a wide variety of unique and interesting electronic properties are available for exploitation. Incorporating these materials into many-layered heterostructures allows an almost infinite combination of structures exploiting their unique properties, thereby enabling a large variety of electronic devices to be produced, connected and directly integrated. In this way, only one family of materials will be required for complete IC fabrication.

In order to produce 2D material based ICs, successful fabrication of a wide variety of working devices must be demonstrated using a simple technique. Inkjet printing provides an attractive method to tackle this problem: it allows maximum design flexibility, and it does not require expensive masks to be made or time-consuming resist process steps, as in traditional electronic fabrication. This allows for quick and easy integration of the various components into a fully-printed IC. Furthermore, inkjet printing is an additive and non-contact technique, which produces no material wastage and is compatible with almost any surface of interest, including flexible substrates.[10–12] Finally, in the case of 2D materials, inkjet printing can also be used to make fully printed heterostructures: all inkjet printed transistors[13,14], photodetectors[15] and memories[15] have been demonstrated. However, these devices are usually integrated with traditional components. Therefore, more work needs to be done to demonstrate an all-2D material fully-printed IC. In this framework, it is of fundamental importance to be able to fully inkjet print all components. In particular, passive components, such as capacitors, occur in large numbers within electrical systems, forming the basis of many sensors and circuits, and therefore must demonstrate high reliability and good performance.[16] Establishing a reliable technology for printing of the dielectric layer is also a stepping stone for obtaining high-performing printed Field Effect Transistors (FETs). The simplest way to fabricate an all-2D material capacitor is to use hexagonal boron nitride (hBN) as a dielectric placed between two graphene (Gr) electrodes, by forming the heterostructure $Gr_B$/hBN/$Gr_T$, where $Gr_B$ and $Gr_T$ are the bottom and top electrodes, respectively. Previous attempts to fabricate capacitor devices using hBN as the dielectric have involved non-printable based techniques, such as chemical vapour deposition (CVD), often using evaporated gold as electrodes, as reported in Refs. [17] and [18]. As a result of high material quality and thickness in the few tens of nm, these devices are capable of achieving large capacitance; however, the methods are expensive and time-consuming. In addition, problems have been reported regarding residual polymer after film transfer[19] and leakage through the dielectric.[18] Solution based techniques have also been investigated for the fabrication of hBN-based capacitors: from hBN membranes made by vacuum filtration[20] to screen-printed hBN.[21] Both methods offer a low-cost approach for producing dielectric films, yet the processes offer poor design flexibility, high waste amounts, and generate thick films, thereby lowering the capacitance. Capacitor devices with inkjet-printed graphene electrodes and a spray-coated hBN dielectric were demonstrated in Ref. [22]. Whilst spray-coating produces thinner films compared with vacuum filtered membranes and screen printing, materials wastage remains high and device design is limited. Layer-by-layer (LBL) assembly of hBN films using complementary polymers allows for precise control of the film thickness and leads to ultrathin dielectrics, which demonstrate high capacitance.[23] However, the annealing temperatures required (600 °C) are too high for many flexible substrates, which often consist of polymeric films that would melt at such a temperature.[23] Notably, Ref. [13] showed an attempt to produce fully inkjet-printed hBN capacitor devices using Ag for the electrodes, obtaining a capacitance of 8.7 nF cm$^{-2}$ for an hBN thickness of 1.2 μm.[13]

In this work, we use water-based and biocompatible Gr and hBN inks to fabricate all-2D material and inkjet-printed capacitors. All inks have been produced using the same process and with the same ink solvent formulation already reported in our previous study,[15] where we also demonstrated the ink's biocompatibility. In Ref. 15 the inks were used to fully print heterostructure-based devices, in particular photodetectors and read–only memories. In this work, we aim at demonstrating a different fully printed heterostructure-based device, exploiting the properties of the hBN ink, *i.e.* a capacitor. Note that Ref. 15 did not show any characterization of the hBN inks and no device containing hBN. Here we provide a full characterisation of the dielectric properties of hBN: we measured an average breakdown field of 1.93 ± 0.3 MV cm$^{-1}$ for printed hBN films of ~3 μm thickness. This value is comparable to the breakdown voltage of many dielectric polymers. More than 100 devices have been fabricated and tested. The fabrication yield depends on the overlapped area and thickness of the hBN, but on average we found a yield of 62%. The Gr$_B$/hBN/Gr$_T$ capacitor devices demonstrate a capacitance of the orders of nF cm$^{-2}$ for a thickness of ~3 μm and negligible leakage currents. Fully printed capacitors with sub-μm hBN layer thicknesses have also been demonstrated. We used the capacitors to demonstrate a fully printed low-pass filter, made of 2D materials only, and a graphene-based field effect transistor.

As one of the fundamental circuitry components, capacitors can be implemented within electronic systems across a number of fields, including wearable or epidermic electronic applications. The biocompatibility of the constituent 2D inks[15] is therefore an attractive feature for devices, which could be used inside the body for *in situ* monitoring[24,25] or for wearable electronics, such as tattoo-based sensors.[26–28]

**RESULTS AND DISCUSSION**

Water-based and biocompatible graphene and boron nitride inks were prepared using liquid-phase exfoliation according to the protocol developed in our previous work.[15] The solvent formulation was specifically engineered to have the correct rheological properties for inkjet printing and to minimise remixing at the interface, allowing fabrication of heterostructures using only inkjet printing. The vertically-stacked $Gr_B/hBN/Gr_T$ capacitors were fabricated by first printing the graphene ink (60 printed passes, at a concentration of ~2 mg ml$^{-1}$) on to a pre-cleaned glass substrate (Fisher Scientific Clear Glass Slides, 1.0-1.2 mm thickness) to produce a ~0.15 mm by ~3 mm graphene rectangle, which acts as the bottom electrode ($Gr_B$). A glass substrate was chosen to avoid any parasitic capacitive effects that might be seen with $SiO_2$. An hBN rectangle, with a width of ~0.5 mm and length in the range of 1-2 mm, was printed across the $Gr_B$ using 80 printing passes of hBN ink at a concentration of ~2 mg ml$^{-1}$. Finally, the top graphene electrode ($Gr_T$), with length ~3 mm, was printed in the centre of the hBN rectangle, perpendicular to the $Gr_B$. In this work, the width of the $Gr_T$ ($\Delta x_T$) was varied incrementally between ~0.15 mm and ~1.28 mm, in order to investigate how the capacitance is changing with increasing overlapping area (A) between the graphene electrodes. The thicknesses of the deposited graphene and hBN films were measured using a profilometer (Supporting Information, Section 'Film Thickness Characterisation'), showing that for 60 printed passes of graphene, the electrode thickness is ~300 nm and for 80 printed passes of hBN, the dielectric film thickness is ~3 μm, with a roughness of ~700 nm.

**Figure 1a** shows a schematic diagram of a $Gr_B/hBN/Gr_T$ capacitor device. Figure 1b shows a magnified image of the active overlap area between the two graphene electrodes. One can observe the uniformity in droplet deposition and the good separation between the $Gr_B$, hBN and $Gr_T$ regions of the heterostructure *i.e.* there is no material intermixing at the interfaces. Re-mixing at the interface is prevented through inclusion of a shear-thinning biocompatible binder within the ink solvent formulation.[15] The shear-thinning nature of the binder ensures that, once deposited onto the substrate, the viscosity of the ink substantially increases, such that mixing between different 2D material inks is minimised within the heterostructure device.[15] An optical picture of 14 devices printed on glass is shown in Figure 1c. Figure 1d, 1e and 1f show optical images of representative devices with increasing A, ranging from 0.0361 to 0.192 mm$^2$.

The impedance of the printed $Gr_B/hBN/Gr_T$ capacitors has been characterised by means of a precision LCR meter (HP 4284A), working in the parallel $R_p$-$C_p$ circuit mode, for which the modulus of the impedance, |Z|, is given by:[20,29]

$$|Z| = \frac{R_p}{\sqrt{1 + \omega^2 R_p{}^2 C_p{}^2}} = \frac{\sqrt{(R_{ESR} + R_{LEAK} + \omega^2 C^2 R_{ESR} R_{LEAK}^2)^2 + (\omega C R_{LEAK}^2)^2}}{1 + (\omega C R_{LEAK})^2}$$

(1)

|Z| can also be expressed considering the equivalent circuit as in the inset of **Figure 2a**, where ω is the angular frequency, $R_{ESR}$ is the series resistance, C is the capacitance and $R_{LEAK}$ is a figure of merit of the leakage through the hBN dielectric. Figure 2a shows |Z| for a capacitor with C = 6 pF (measured at 1 kHz) up to 1 MHz (red circles) fitted *via* the $R_{ESR}$, $R_{LEAK}$ and C parameters. As can be seen, $R_{LEAK}$ has a value >1 GΩ (negligible losses) and $R_{ESR}$ has a value of 90 kΩ, with ideal capacitor behaviour up to $10^5$ Hz. An $R_{ESR}$ value of 90 kΩ corresponds well to the measured resistance values normally achieved for Gr electrodes deposited *via*

inkjet printing using our water-based inks.[15] This value can be easily adjusted by changing the thickness and the width of the graphene electrodes.

Figure 2b and 2c show the measurements performed with an HP4284A LCR meter *i.e.* $C_p$ and $R_p$ as a function of frequency, in the frequency range between 1 kHz and 1 MHz, for devices with increasing A (*i.e.* obtained by increasing $\Delta x_T$ from 500 μm to 800 μm, with constant dielectric thickness of ~3 μm ($\Delta x_B$ = 170 μm)). One can observe the decrease in both capacitance ($C_p$) and parallel resistance ($R_p$) with increasing frequency, in agreement with Refs. [18,21,22]. Indeed, as expected, the parameters of the equivalent parallel circuit representation ($R_p$ and $C_p$), depend on the frequency.[30] Figure 2b also shows that the $C_p$ for a given frequency increases with increasing overlap area, as expected in a capacitor realized with standard fabrication techniques. Similar results were reported in Ref. [22]: the larger the area, the larger the capacitance. Figure 2d shows the $C_p$ plotted *vs.* frequency for two and three capacitor devices for a fixed A of 0.036 mm$^2$, connected in parallel. The obtained values at a given frequency scale proportionally to the number of devices connected. The ability to connect capacitors in parallel is crucial for achieving large enough capacitance values in ICs. **Figure S9** shows the capacitance values obtained for a set of devices with constant overlap area (A = 0.075 mm$^2$) and variable thickness. As in traditional capacitors, the capacitance decreases for increasing dielectric thickness. In order to obtain a large dataset of capacitance values, the active capacitor overlap area was varied by printing a different width of top graphene electrode for each set of devices, keeping all other parameters constant. **Figure 3** shows the average capacitance value (measured at 1 kHz) plotted as a function of A/t, for each set of devices printed, on log-log scale. The data fitting (black line, Figure 3) highlights a linear increase in capacitance with increasing A/t, as expected using the parallel plate capacitor model.[30,31] Following this model, the capacitance is given by:[31]

$$C = \varepsilon_0 \varepsilon_r \frac{A}{t} \tag{2}$$

where $\varepsilon_0$ is the electric constant of vacuum and $\varepsilon_r$ is the relative dielectric constant. Thus, the y-axes offset of the C *vs*. A/t data on the log-log scale (Figure 3) gives $\varepsilon_0\varepsilon_r$. From the data in Figure 3, we calculated a dielectric constant of 6.1, which is in agreement with that reported for CVD hBN and inkjet printed Ag/hBN/Ag capacitors.[13,17] However, a strong discrepancy is observed with the points from Ref. [22], which have the same slope, but different y-axes offset, and thus a different dielectric constant. Indeed, by performing a comparison with all data in literature, one can see a large spread of the dielectric constant values reported for hBN. We can divide the data into 3 groups: those with a dielectric constant of 2-4;[22,32–37] others, including our work, reporting a dielectric constant between 5 and 12[13,17] and finally, one group reporting a dielectric constant of >200.[18] The data are summarized in Table S1. The reason why our dielectric constant is higher than that of bulk hBN could be attributed to the chemical composition (additives, residual solvent, *etc.*) and the morphology of the deposited films (*e.g.* a high interfacial roughness increases the real capacitor area, compared to the geometric area). Previous works have measured the dielectric constant of hBN membranes produced using inks based on *N*-Methyl-2-pyrrolidone (NMP)[22] or mixed solvents[38], while our inks, although based on water, also contain stabilisers and additives.[15] Thus, the different composition of the starting ink, including residual water, could reflect in a different dielectric constant. The surface roughness measured for our capacitors (Supporting Information, Section 'Topography Characterisation') is relatively small compared to the area of the film, and thus the difference between geometric and real area of the capacitor should be negligible. In addition, different deposition techniques give rise to different film morphologies (Supporting Information, Section 'Characterisation of hBN Dielectric Properties'), which can affect the orientation of the flakes and the porosity, which in turn will determine the dielectric constant. For example, we have fabricated an hBN membrane by vacuum filtration with the same printable ink used in this work, and measured a dielectric constant of ~2 (Section 'Characterisation of hBN Dielectric Properties', Figure S8).

The printed hBN layer demonstrates a good dielectric strength of 1.93 MV cm$^{-1}$ and an average breakdown voltage of ~600 V across seven devices ($\Delta x_B$ = 190 μm, $\Delta x_T$ = 190 μm, t = 3.1 μm). Breakdown voltages for each of these seven devices are shown in Table S2. The dielectric strength is comparable to that of hBN measured previously in our group through fabrication of a vacuum filtered membrane (2.5 MV cm$^{-1}$, in vacuum)[38] and in agreement with the range reported in literature for hBN (1.5-2.5 MV cm$^{-1}$).[13,32] We can also observe that the dielectric strength of the inkjet printed hBN films is comparable to that of well-known dielectric materials (Table S3).[39] In particular, dielectric polymers have been widely used in electronic devices as a result of their amenability to solution-processing and low cost patterning techniques. A wide variety of chemical structures are available and careful control of the polymerisation reaction conditions enables the material characteristics and dielectric properties to be tuned.[40] As the dielectric strength and dielectric constant of our hBN films are comparable with that of dielectric polymers, hBN shows strong potential for use in organic electronics. We remark that hBN requires relatively high thickness compared to polymers; on the other side, many dielectric polymers (in particular, high-k polymers) are not suitable for inkjet printing due to their high viscosity and solubility in a limited number of solvents, which often are not suitable for inkjet printing (*e.g.* anisole, n-butyl acetate).

In order to study the yield of Gr$_B$/hBN/Gr$_T$ capacitor devices fabricated through inkjet-printing, multiple devices (between 2 and 14) were printed for each selected value of overlap area. Devices showing a linear I-V characteristic *i.e.* behaved as a short circuit, were classified as non-functional. Short circuits are a result of occasional inconsistencies in the deposited film which occur when the inkjet printer nozzle becomes blocked. Additionally, in some instances, the leakage through the hBN dielectric was considered to be too high for the device to be truly classified as a functional capacitor. To be considered functional, we selected only devices satisfying the following condition: $R_{LEAK} > 10/(\omega C)$. This condition ensures that the capacitive part of the impedance is prevalent over the resistive part. The corresponding data regarding the number of devices printed and the number of those devices that were functional across the various A/t ratios can be found in the supporting information (Table S4). The yield is dependent on the overlap area, but overall an average yield of 62% has been found.

Finally, we exploit the use of the printed capacitors within devices. We have fully printed on a glass substrate a simple RC low-pass filter, composed by a graphene resistance, R = 17 MΩ and a capacitance, C = 7.6 pF, as shown in the inset of **Figure 4a**. In particular, the resistor has been fabricated by means of graphene lines (5 printed passes), while the capacitor is composed of graphene top and bottom electrodes (60 printed passes), embedding 70 printed passes of hBN dielectric. Figure 4a shows that the measured frequency response of the filter is in very good agreement with theoretical results, with a cut-off frequency of approximately 1.2 kHz (*i.e.*, theoretical cut-off frequency is $f = (2\pi RC)^{-1} = 1.232$ kHz).

In order to demonstrate that the developed technology can be applied also in Field Effect Transistors and on flexible substrates, we have fabricated a Graphene-based Field Effect Transistor (GFET) printed on paper. In particular, paper is selected because it is one of the cheapest flexible substrates, which is also recyclable and foldable, and allows achievement of the best electrical performance with our inks.[15] The inset of Figure 4b shows the GFET layout, where source, drain and gate electrodes are printed with Ag ink, while the channel and the dielectric have been printed by means of graphene and hBN inks, respectively. The sketch of the longitudinal cross section is shown in the inset of Figure 4c. The length and the width of the channel are 70 μm and 500 μm, respectively. 25 printed passes of hBN were deposited and the substrate used was paper (see Materials and Methods).

The transfer and the output characteristics are shown in Figure 4b and Figure 4c. As can be seen, the gate is able to modulate the current in the channel, while keeping the gate leakage always lower than 1 nA, in the whole range of applied gate voltages.

**CONCLUSIONS**

Herein, we have shown all-2D material inkjet printed $Gr_B$/hBN/$Gr_T$ capacitors as passive components for IC. The achievable capacitance can be easily tuned through adjustment of the device active area and by connecting multiple devices in parallel or series. For an hBN dielectric thickness of ~3 μm, an average areal capacitance of 2.0 ± 0.3 nF cm$^{-2}$ was measured. The inkjet printed hBN has a dielectric constant of 6.1 ± 1.7 and breakdown strength of 1.9 ± 0.3 MV cm$^{-1}$. The capacitor component can be easily connected to others, *e.g.* other printed capacitors and graphene resistors, to form simple all-2D material circuits. More complex circuits can be made by increasing the library of printed components, such as diodes and high-mobility transistors.

**MATERIALS AND METHODS**

*Ink Preparation*: Bulk graphite (purchased from Graphexel or Sigma-Aldrich, 99.5% grade) and bulk boron nitride (purchased from Sigma-Aldrich, >1 μm, 98% grade) powders were used to prepare the inks. The bulk powders were dispersed in de-ionised water (resistivity 18.2 MΩ cm$^{-1}$) at a concentration of 3 mg mL$^{-1}$ and 1-pyrenesulphonic acid sodium salt (PS1), purchased from Sigma-Aldrich, purity ≥ 97%, was added at a concentration of 1 mg mL$^{-1}$. The graphite and boron nitride dispersions were sonicated for 72 h and 120 h respectively using a 300 W Hilsonic HS 1900/Hilsonic FMG 600 bath sonicator at 20 °C. The resultant dispersions were centrifuged at 3500 rpm (g factor = 903) for 20 minutes at 20 °C using a Sigma 1-14K refrigerated centrifuge in order to separate out and discard the residual bulk, non-exfoliated flakes. The remaining supernatant, now containing the correct flake size and monolayer percentage, was centrifuged twice to remove excess PS1 from the dispersion. After washing, the precipitate was re-dispersed in the printing solvent, made as described in Ref. [15].

The concentration of the resultant inks were assessed using a Varian Cary 5000 UV-Vis spectrometer and the Beer-Lambert law, with extinction coefficients of 2460 (at 660 nm) and 1000 L g$^{-1}$ m$^{-1}$ (at 550 nm) for graphene[41] and hBN[42], respectively. The inks used for printing were diluted to a concentration of 2 mg mL$^{-1}$.

*Printing*: A Dimatix DMP-2800 inkjet printer (purchased from Fujifilm Dimatix) was used to print the Gr$_B$/hBN/Gr$_T$ capacitor devices onto glass microscope slides (Fisher Scientific Clear Glass Slides, 1.0-1.2 mm thickness) using a 16-nozzle cartridge with 23 μm nozzle diameter and typical droplet volume of 10 pL. The printer platen was heated to 60 °C and a drop spacing of 40 μm was utilised, as this gives the smaller sheet resistance for the printed film.[15] For both the top and bottom graphene electrodes, 60 printing passes were deposited and for the hBN dielectric layer 80 printing passes were deposited. A step-by-step annealing procedure was employed, in which the devices were annealed under vacuum for 2 hours at a temperature of 150 °C following the deposition of each layer in the Gr$_B$/hBN/Gr$_T$ heterostructure stack.

The RC low-pass filter, also printed onto a glass slide substrate, is a combination of a Gr$_B$/hBN/Gr$_T$ capacitor device with a printed graphene resistor. For the capacitor, the Gr$_B$, hBN and Gr$_T$ films were deposited using 60, 70 and 60 printed passes of Gr, hBN and Gr ink respectively. For the resistor, a serpentine layout was produced using 5 printed passes of Gr ink.

A Gr-based field effect transistor was also printed onto a paper substrate (PEL P60, from Printed Electronics Limited) using silver ink (1 printed pass, Sigma-Aldrich) for the source, drain and gate electrodes. Graphene ink was used to deposit the channel (25 printed passes) and hBN ink (25 printed passes) to fabricate the dielectric film. The channel has a length of 70 μm, a width of 500 μm and a resistance of 10 kΩ.

*Characterisation*: Line scans were taken to measure the thickness of the printed features using a Bruker Dektak XT Stylus Profiler (stylus radius of 12.5 μm, stylus force of 3 mg, scan speed of 100 μm s$^{-1}$, scan resolution of 0.33 μm). Gwyddion SPM analysis software was used to generate RMS roughness and effective surface area values from the 2D profilometry map data. In order to further study the surface topology of the printed films, AFM images were taken using a Bruker MultiMode 8 in PeakForce QNM mode with ScanAsyst-Air probes. Cross-sectional SEM images were taken using a Zeiss Sigma HV instrument.

*Capacitance Measurements*: I-V characteristics for the $Gr_B$/hBN/$Gr_T$ capacitor devices were acquired using a Keithley 4200 Semiconductor Characterization System parameter analyser. Capacitance data were collected in vacuum in the frequency range 1 kHz to 1 MHz using an HP 4284A Precision LCR meter with short-circuit and open-circuit correction (bias voltage, $V_b = 0$ V and measurement voltage, $V_m = 1$ V). Fitting was performed using an RC parallel circuit model.[43] The capacitor electrode has been connected to the measurement system through the tip of the probe station, touching a silver pad previously defined.

*Dielectric constant measurements*: The hBN ink was assessed for use as a dielectric material through measurement of the dielectric constant. To measure the dielectric constant, a membrane was produced through vacuum filtration using the printable hBN ink. A total of 20 mg of hBN material was deposited, giving a membrane thickness of 170 μm. Gold contacts were evaporated onto either side of the membrane and a capacitance measurement was made, allowing the dielectric constant to be calculated using Equation 2 (Main Text).

The breakdown voltage and breakdown field were measured for 7 capacitor devices (electrode overlap area of 0.036 mm$^2$, capacitance ~2 pF) using the Keithley 4200 Semiconductor Characterization System parameter analyser.

ASSOCIATED CONTENT

Conflict of Interest: The authors declare no competing financial interest.

AUTHOR INFORMATION

Corresponding Author E-mail: cinzia.casiraghi@manchester.ac.uk

ACKNOWLEDGMENTS

This work is partially supported by the European Research Council (ERC) under the European Union's Horizon 2020 research and innovation programme under grant agreement No 648417 and No 770047, and by the project HETERO2D and the Graphene core 2 under grant No 785219. C. Casiraghi and K. Novoselov acknowledge the Grand Challenge EPSRC grant EP/N010345/1. D. McManus acknowledges funding from the EPSRC in the framework of the CDT Graphene NOWNANO. R. Worsley acknowledges M. Turner for useful discussions.

**Supporting Information Available**: Hexagonal boron nitride ink characterisation; film thickness and topography characterization; hBN dielectric constant, breakdown voltage, dielectric strength measurements; statistics and effect of hBN thickness. This material is available free of charge *via* the Internet at http://pubs.acs.org.

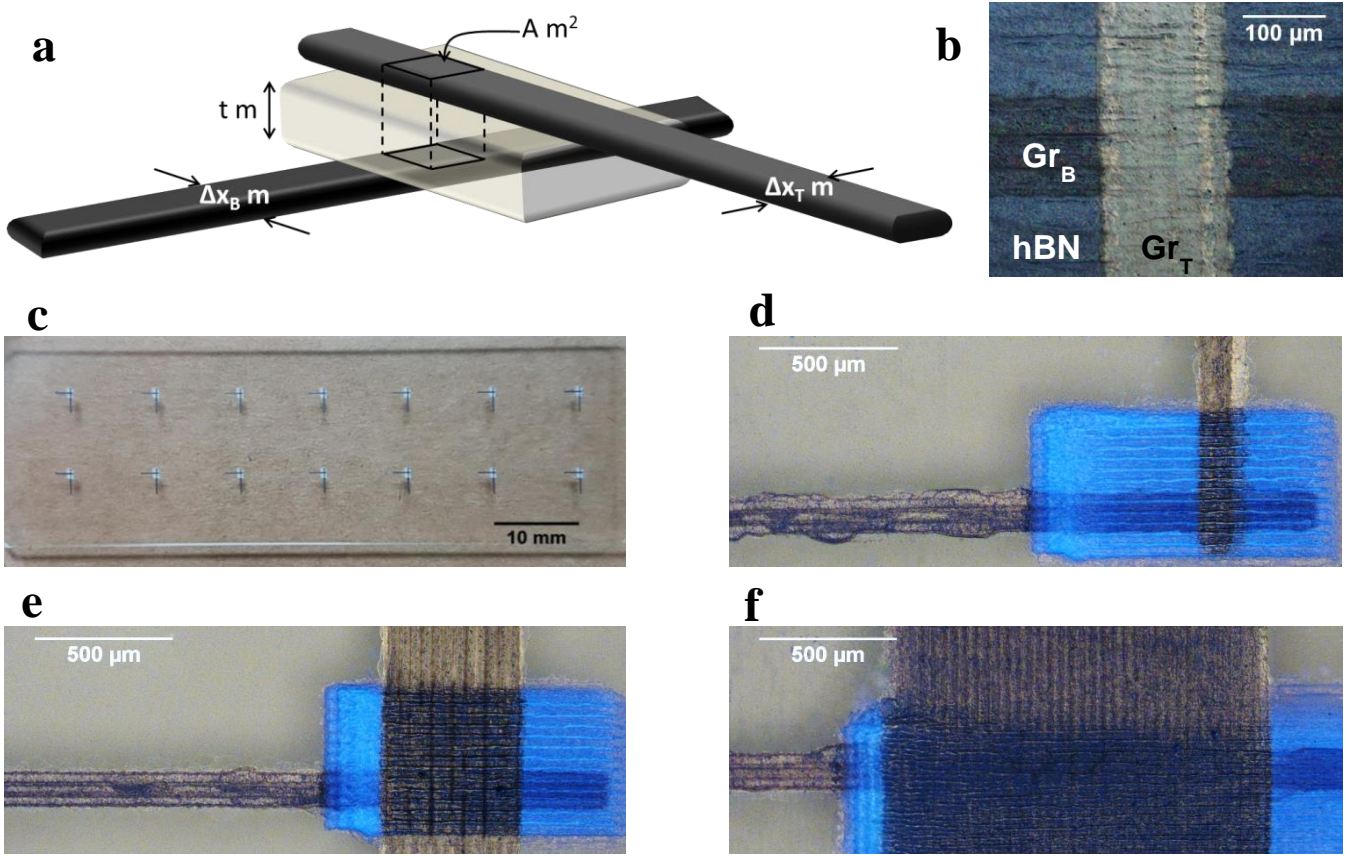

**Figure 1. a)** Schematic diagram of $Gr_B$/hBN/$Gr_T$ inkjet printed capacitor device, defining the width of the bottom Gr electrode ($\Delta x_B$), the width of the top Gr electrode ($\Delta x_T$) and the electrode overlap area (A). b) Magnified image of the overlap region between the two graphene electrodes, clearly showing the separation between the layers of the heterostructure. c) Photograph of 14 capacitor devices printed onto glass. d, e, f) Optical images of representative small, medium and large area capacitor devices, respectively

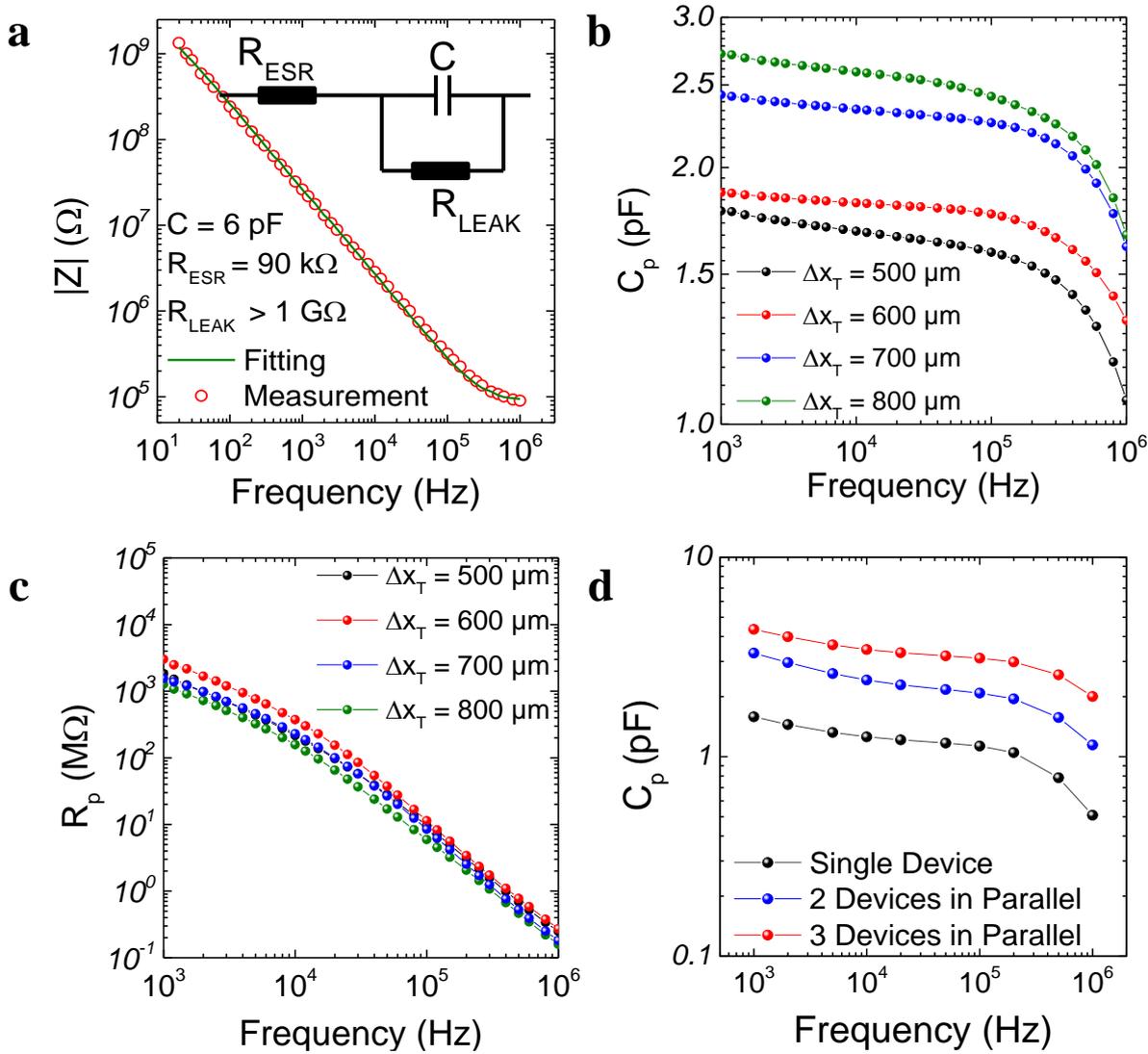

**Figure 2. a** Impedance modulus values for a representative $Gr_B/hBN/Gr_T$ capacitor device plotted as a function of frequency. The fitting values shown correspond to a 'leaky capacitor with equivalent series resistance (ESR)' model, inset. **b)** Representative measured capacitance values plotted as a function of frequency for inkjet printed $Gr_B/hBN/Gr_T$ capacitors with bottom graphene electrode width $\Delta x_B = 170$ μm, hBN thickness $t = 3.1$ μm, and the width of the top graphene electrode, $\Delta x_T$, is varied between 500 μm and 800 μm. **c)** Representative resistance values plotted as a function of frequency for inkjet printed $Gr_B/hBN/Gr_T$ capacitors where $\Delta x_T$ is varied between 500 μm and 800 μm. **d)** Measured capacitance values plotted as a function of frequency for inkjet printed $Gr_B/hBN/Gr_T$ capacitors connected in parallel.

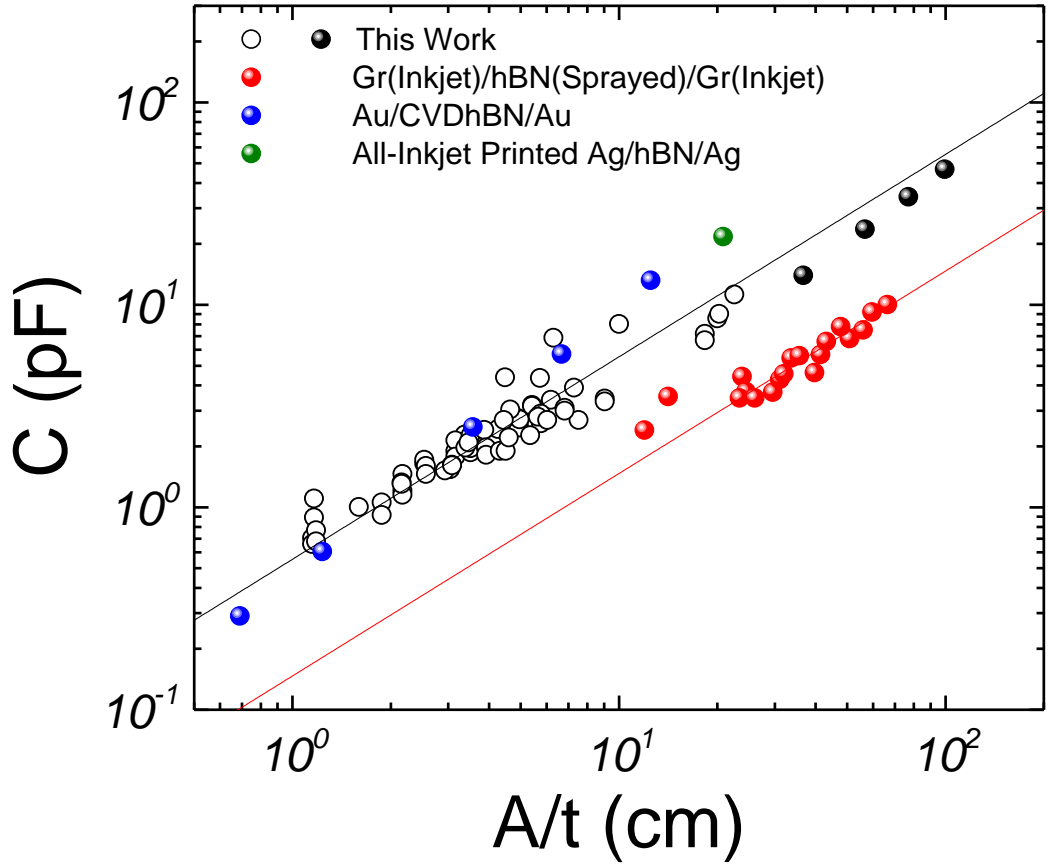

**Figure 3.** Capacitance plotted as a function of the area to thickness (A/t) ratio for the all-inkjet-printed $Gr_B/hBN/Gr_T$ capacitors in this work (open circles – single capacitance; black circles – capacitances in parallel), Au/CVD-hBN/Au devices with values taken at 2 kHz (blue circles)[17], Gr/hBN/Gr devices in which the hBN has been spray-coated (red circles)[22] and all-inkjet-printed Ag/BN/Ag devices (black circle)[13]. The black line is a linear fit to our experimental data. The red line is a linear fit to the experimental data from Ref. 22.

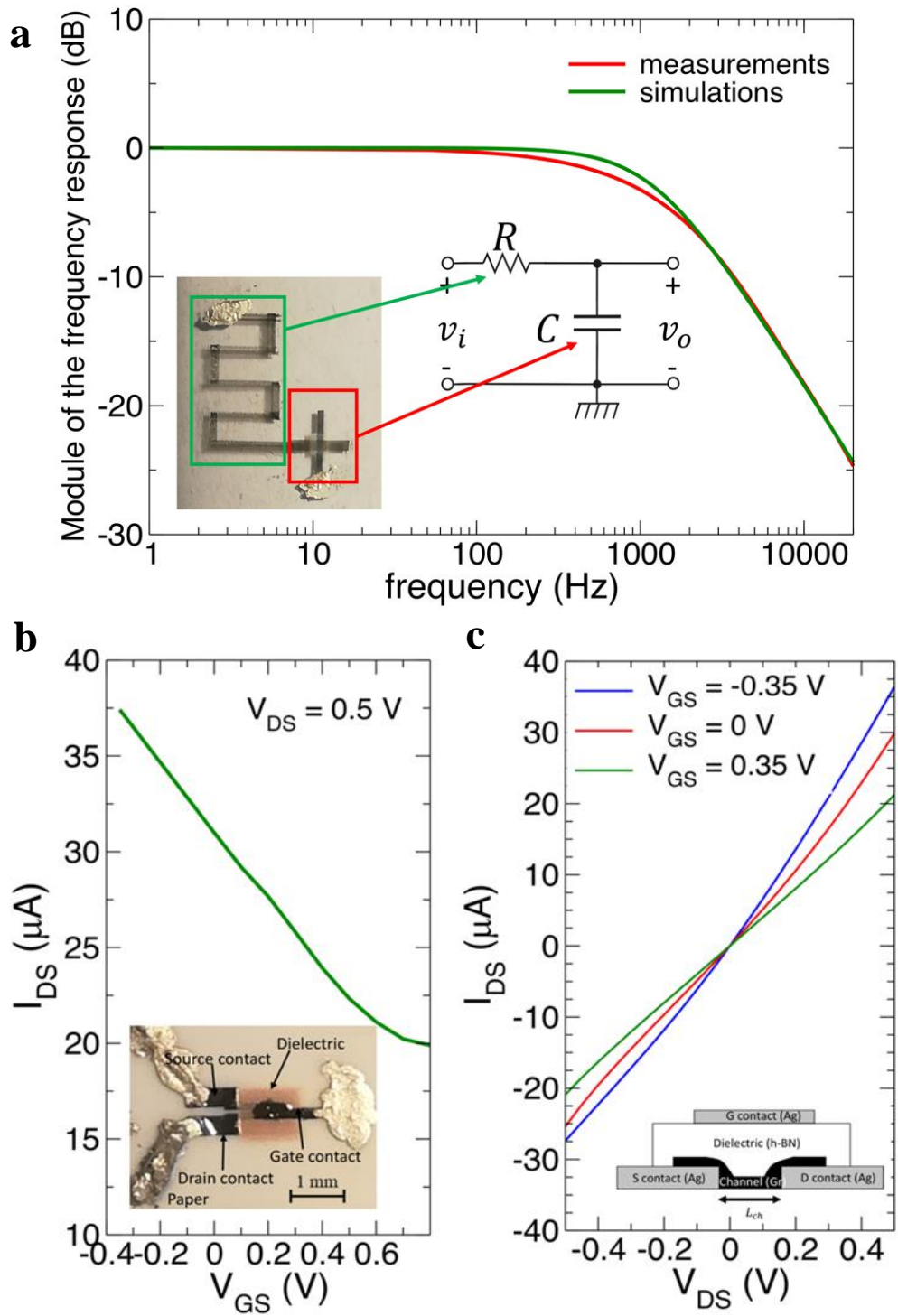

**Figure 4. Demonstrator Devices. a)** Module of the frequency response of a first-order low-pass RC filter. Both experimental (red) and theoretical (green) results are shown. A photograph of the printed circuit is shown in the inset. **b, c)** Transfer and output characteristics of the fabricated GFET. The layout and the longitudinal cross-section of the GFET are shown in the insets.